# Ultrafast pump-probe phase-randomized tomography

Filippo Glerean[1,2,3], Enrico Maria Rigoni[1,2], Giacomo Jarc[1,2,5], Shahla Yasmin Mathengattil[1,2], Angela Montanaro[1,2,5], Francesca Giusti[1,2], Matteo Mitrano[3], Fabio Benatti[1,2,4], and Daniele Fausti[1,2,5,*]

[1] Dipartimento di Fisica, Università degli Studi di Trieste, Trieste I-34127, Italy
[2] Sincrotrone Trieste S.C.p.A., Basovizza I-34149, Italy
[3] Department of Physics, Harvard University, Cambridge, MA-02138, USA
[4] Istituto Nazionale di Fisica Nucleare, Sezione di Trieste, Trieste I-34014, Italy
[5] Department of Physics, University of Erlangen-Nürnberg, 91058 Erlangen, Germany
[*] Correspondence: daniele.fausti@fau.de

**Measuring fluctuations in matter's low energy excitations is the key to unveil the nature of the non-equilibrium response of materials. A promising outlook in this respect is offered by spectroscopic methods that address matter fluctuations by exploiting the statistical nature of light-matter interactions with weak few-photon probes. Here we report the first implementation of ultrafast phase randomized tomography, combining pump-probe experiments with quantum optical state tomography, to measure the ultrafast non-equilibrium dynamics in complex materials. Our approach utilizes a time-resolved multimode heterodyne detection scheme with phase-randomized coherent ultrashort laser pulses, overcoming the limitations of phase-stable configurations and enabling a robust reconstruction of the statistical distribution of phase-averaged optical observables. This methodology is validated by measuring the coherent phonon response in α-quartz. By tracking the dynamics of the shot-noise limited photon number distribution of few-photon probes with ultrafast resolution, our results set an upper limit to the non-classical features of phononic state in α-quartz and provide a pathway to access non-equilibrium quantum fluctuations in more complex quantum materials.**

Fluctuations are a fundamental feature of quantum systems and revealing them is a key challenge in understanding some of the most debated exotic states in complex quantum materials[1] and in designing new quantum devices[2]. Quantum phenomena like superposition, entanglement and vacuum fluctuations have an inherently statistical nature, which can lead to intriguing macroscopic effects in quantum materials when the quantum correlations survive thermal decoherence. Ultrafast photoexcitation has recently emerged as a powerful means to control and induce new coherent phenomena, like light-induced superconductivity[3,4,5,6], light-induced ferroelectricity[7,8] and vibrational light-induced transparency[9], which are otherwise not accessible at the thermodynamic equilibrium. Harnessing these non-equilibrium states requires understanding how the fluctuations of the relevant electronic, vibrational or magnetic degrees of freedom are modifying the natural thermal evolution of the system.

Treating the light-matter coupling fully at the quantum level, beyond semiclassical approximations, opens new spectroscopic opportunities[10] to access the fluctuations in materials. The strategy is to investigate the statistical degrees of freedom of matter leveraging on the knowledge of the statistical properties of quantum light developed in quantum optics. The quantum optical properties are for instance playing a role when considering the ultrafast electron dynamics driven by intense light. Although the strong electric field has always been so far considered classical, the quantum statistical distribution of the light has been proposed to induce the emission of High Harmonic Generation radiation[11,12,13,14] or electrons[15] with new properties.

We explore the quantum character of ultrafast light-matter interaction with a different perspective. Rather than studying the effects of quantum light as input, we investigate how materials can modify the quantum statistical properties of the output light and propose to study the intrinsic quantum fluctuations of the system by imprinting them into the statistical properties of light. Quantum spectroscopies have been successful in studying quantum fluctuations at equilibrium[16,17,18,19], but their application to ultrafast non-equilibrium phenomena is so far limited to theoretical efforts[10,20] and a few experimental attempts[21,22,23], because it is technically challenging to adapt standard pump-probe experiments to reliably detect the quantum statistical response.

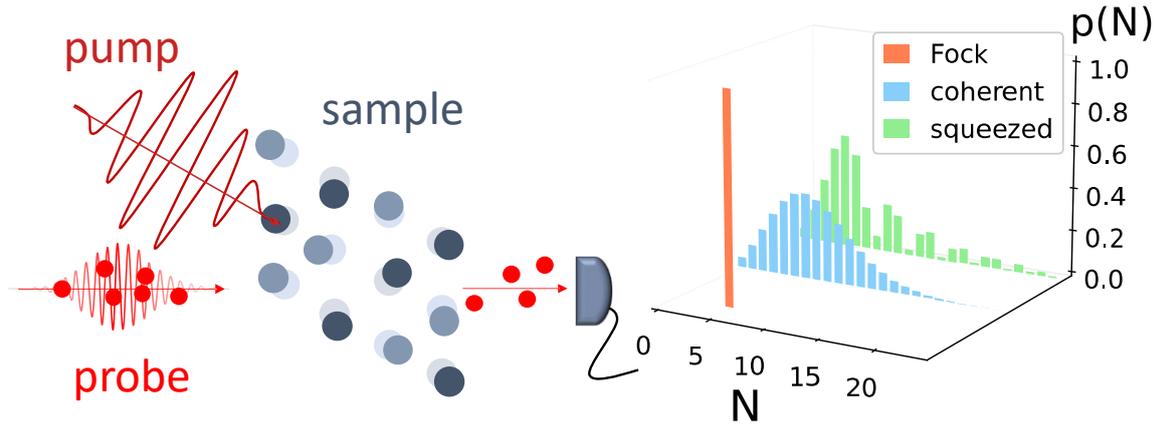

**Figure 1: Ultrafast pump-probe spectroscopy of fluctuations measuring the quantum optical statistics of ultrashort laser pulses.** We investigate the possibility to access the system fluctuations distinguishing the effects that the light-matter interaction produces in the photon number distribution of weak probe pulses, altering the classical coherent state statistics.

In a pump-probe experiment, a first strong pump pulse impulsively drives the system out of equilibrium and a second probe pulse monitors with femtosecond resolution the relaxation dynamics of the sample excitation. Pump-probe experiments usually detect tiny changes in the average optical intensity of classical probe fields, while the quantum properties of light emerge especially in the weak intensity regime, where the photon discretization comes into play. The statistics of weak light pulses with a few photons per pulse is difficult to measure since direct low-photon counting detection schemes are still at a developing stage[24,25] and indirect quantum state reconstruction methods rely on delicate and slow phase-resolved interferometric measurements[23,26].

In this work, we devise an ultrafast quantum spectroscopy method which measures the probe photon number statistics without a phase-stable interferometer, taking advantage of coherent phase-randomized, or phase averaged (PHAV)[27,28] states. Weak PHAV states are employed as realistic single-photon sources[29,30,31], which are useful to implement decoy states in quantum communication protocols[32], random number generation[33] and reveal quantum interference[34]. We develop a phase-randomized heterodyne interferential scheme for measuring the phase-averaged optical quadrature[35,36], which exploits the intrinsic Carrier-Envelope Phase (CEP) instability of the pulsed laser source. Thanks to phase randomization, we uniformly sample the optical phase space of the probe field and obtain the technical advantage of not being affected by phase stability issues. It is not necessary to measure the phase-resolved mean-value oscillation of the optical quadrature field, but we collect the phase-averaged quadrature distribution and reconstruct with tomography the full photon number distribution of the probe state. We highlight that, since the natural CEP fluctuations are perfectly uniformly distributed and uncorrelated, our method is more reliable and efficient than any phase manipulation protocol (see supplementary material for detailed characterization and discussion).

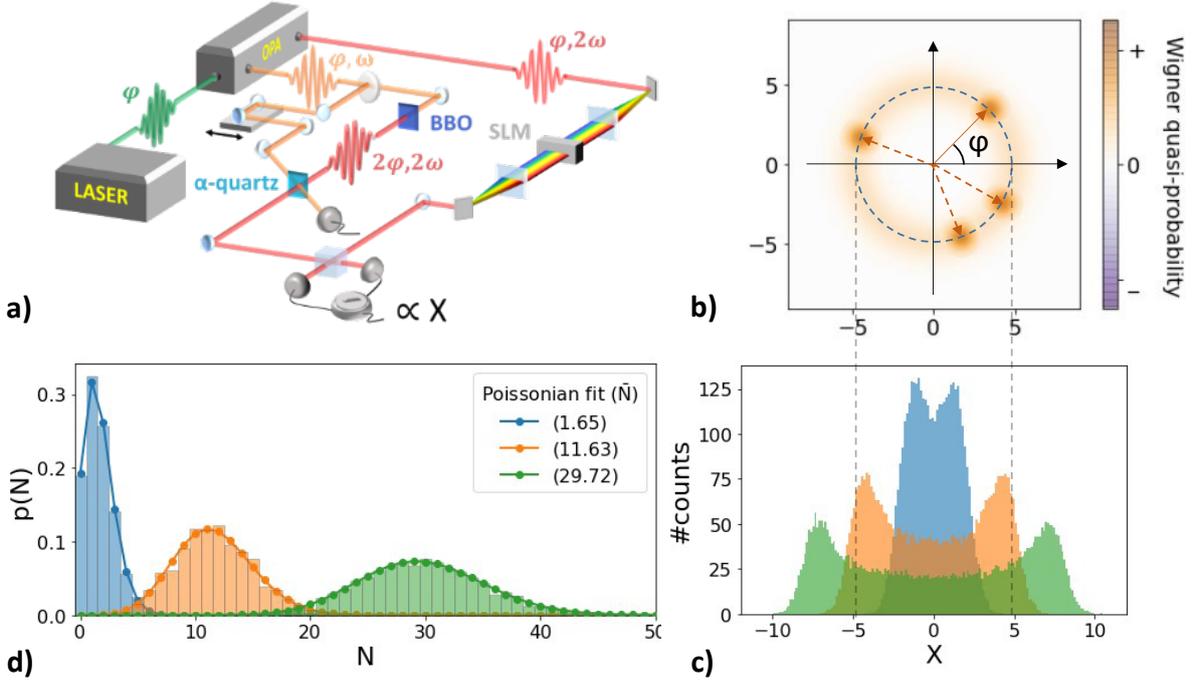

**Figure 2: Reconstruction of the photon number distribution with a phase-randomized pump-probe experiment. a)** Experimental scheme. The signal and idler outputs of an Optical Parametric Amplifier (OPA), such that $\omega_{SIG} = 2\omega_{IDL}$, are used in combination with a Second Harmonic Generation process to setup a phase-averaged pump-probe heterodyne detection sensitive to the random laser CEP (details in text). **b)** Wigner distribution of the phase-averaged coherent state resulting from the randomization of the CEP dependent LO-probe phase. **c)** The detection output is the distribution of the phase-averaged quadrature. **d)** Applying the tomography procedure to the quadrature data we obtain the probe photon number distribution.

## Phase-randomized ultrafast optical tomography

The experimental setup (represented in Fig. 2a) stems from the combination of a pump-probe scheme with optical state tomography. It is an evolution of a multimode heterodyne interferometer[26], optimized for the study of phase-averaged observables of the weak optical probe. The ultrashort pulses provided by the laser source are coherent states. The measurement of the quantum statistics of weak coherent states relies on continuous variable analysis performed through optical tomography[37]. The coherent state is represented in the optical phase space (Fig. 2b) as a minimum uncertainty Wigner distribution, characterized by an amplitude, $\alpha$, and phase, $\varphi$, in analogy with a classical field. The Wigner function $W$ is associated with the quantum optical state and allows us to predict the mean-values of the generic observable $O$ with an integration over the full phase-space

$$\langle O \rangle = 2\pi\hbar \int \int dX dY \, W(X,Y) \, \tilde{O}(X,Y) \tag{1}$$

where $X$ and $Y$ are the two phase-space quadratures and $\tilde{O}$ the Wigner-Weyl transform of the observable $O$. The Wigner-Weyl transform of the operator $O$ is defined as $\tilde{O} = 1/2\pi\hbar \int dx \, \langle X + x/2 | O | X - x/2 \rangle e^{ixY/\hbar}$. The Wigner function $W$ is the Wigner-Weyl transform of the density operator that describes the quantum state. The Wigner distribution can be reconstructed through the tomography algorithm by measuring the generalized quadrature for different phase projections in the optical phase-space as

$$X\phi = \frac{1}{\sqrt{2}}(ae^{i\phi} + a^\dagger e^{-i\phi}) \tag{2}$$

where $a$ and $a^\dagger$ are the ladder operators related to the quantized optical mode. The quadrature is usually measured with a homodyne detection setup, where the intense classical Local Oscillator (LO) field amplifies the weak quantum optical probe and their interference is detected with a balanced detection scheme. The projection phase is referenced as the relative phase $\phi$ between probe and LO beam, which in a conventional phase-stable interferometer is controlled modifying the optical delay between the two. To study phase-averaged observables of a coherent state, such as the photon number distribution, we do not need to resolve the phase-dependent field profile, but we can alternatively measure the statistical distribution of the phase integrated quadrature[35,36,38] (Fig. 2c). In detail, we employ phase-averaged coherent states which have a ring-like Wigner distribution in the optical phase-space as a result of the integration around all the possible phases (Fig. 2b)

$$X_{\text{PHAV}} = \frac{1}{2\pi}\int_0^{2\pi} d\phi \frac{1}{\sqrt{2}}\left(a e^{i\phi} + a^\dagger e^{-i\phi}\right), \qquad (3)$$

which has a distribution whose width depends on the photon number (Fig. 2c). Starting from the PHAV quadrature distribution, we use a Maximum Likelihood algorithm[39,40,41] to calculate the photon number distribution on the Fock space (details on the supplementary). We show in Fig. 2d the agreement between the Poissonian shape and the reconstructed data for probes with different mean-value of photons per pulse. We report that the numerical limitations of the algorithm allow us to calculate the distribution up to 150 photons (see supplementary).

The present discussion is limited to a single frequency component of the photon field. Nevertheless, we note that the ultrashort pulses are multimode. We underline that exploiting the shaping of the LO spectrum we can frequency-resolve the probe response[26] (a characterization of the multimode equilibrium spectrum is reported in the supplementary). Here, we show only the results relative to a narrow frequency band at the center of the probe spectrum, which is representative of the time dependent response observed in all the spectral bandwidth[42].

We generate the PHAV probe pulses for our experiment exploiting the random Carrier-Envelope Phase (CEP)[43] of ultrashort laser pulses. In a conventional homodyne scheme, the probe and LO are split from the same laser beam and the random CEP phase is conveniently cancelled out in the interference between the two. Importantly, we instead preserve the CEP fluctuation to measure a uniformly distributed set of quadrature phases, without the need to resolve the quadrature phase with a stable interferometer. To achieve sensitivity to the random CEP phase, $\varphi_{CEP}$, we take advantage of the Second Harmonic (SH) generation process, which is widely exploited to implement CEP control systems[44]. As depicted in Fig. 2a, the Signal and Idler outputs of an Optical Parametric Amplifier (OPA) are tuned such that the SH of the Idler is resonant to the Signal. The idler beam is split in two: one portion is used as pump and routed through a delay stage, while the other generates the probe beam with a SH generation process. The relative CEP between the two OPA outputs is the same because both the seed white-light generation and the amplification stages are pumped by the SH of the laser fundamental[45]. The probe is generated as SH of the Idler beam on a BBO crystal. Prior to the interaction with the sample, bandpass filters remove the Idler fundamental and neutral filters attenuate the probe to the few-photon regime. The OPA Signal beam is employed as LO. The SH process doubles the CEP and the relative phase between probe and LO results

$$\phi = \varphi_{probe} - \varphi_{LO} = 2\varphi_{CEP} - \varphi_{CEP} = \varphi_{CEP} \qquad (4)$$

which makes the quadrature phase $\phi$ randomized, as the initial laser CEP $\varphi_{CEP}$.

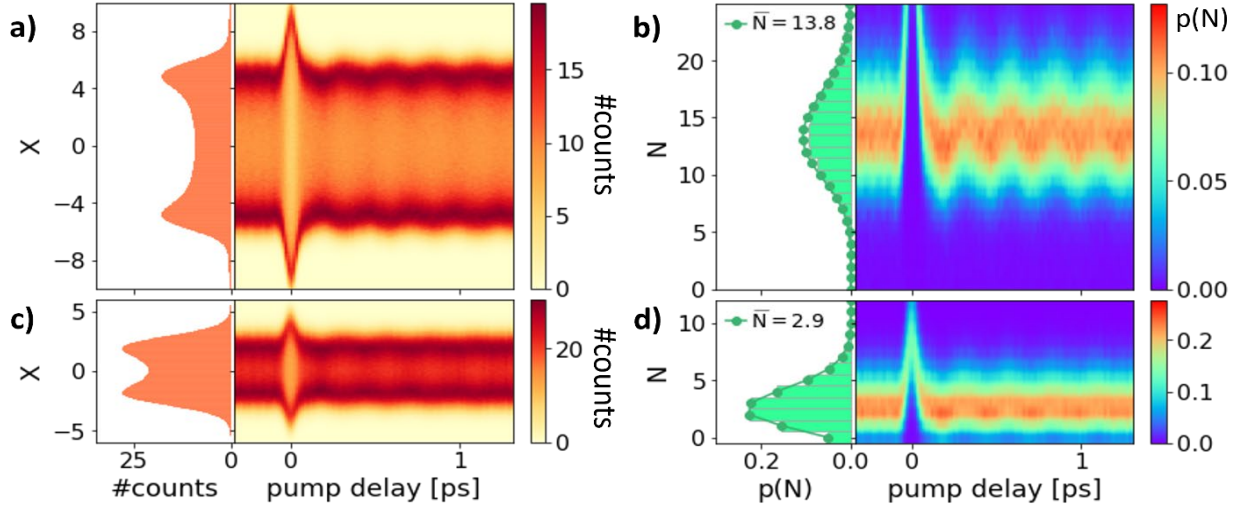

**Figure 3: Pump-probe modulation of the photon number distribution induced by the coherent phonon excitation. a)** Phase-averaged quadrature distribution for a probe pulse with a mean photon number of 13.8 photons per pulse. Left: histogram of the equilibrium phase-averaged quadrature distribution. Right: histogram map describing the time-resolved dynamics of the quadrature distribution. **b)** Applying the tomography algorithm to the quadrature data we study the evolution of the photon number distribution. Equilibrium (left) and time-resolved evolution (right) are showed. We also consider a weaker probe beam with on average 2.9 photons per pulse and report the relative evolution for the quadrature **(c)** and photon number **(d)** distributions. We observe in both cases the presence of coherent oscillations at the phonon frequency.

**Phonon dependent photon number distribution**

We apply the pump-probe PHAV tomography to the study of the non-equilibrium statistical properties of materials. We benchmark the methodology studying the coherent phonon excitation by means of Impulsive Stimulated Raman Scattering (ISRS)[46,47,42] in α-quartz[42,48,49], which is a prototypical example of interaction between electromagnetic fields and vibrational states in matter. A proper selection of the polarization of the pump and probe allows for the selection of the response associated to a phonon mode of specific symmetry (the E-symmetry mode at 4 THz)[50]. We orient the pump at +45° with respect to the probe. We study the probe scattering in the weak residual polarization, which we select with an analyzer orthogonal to the main probe polarization axis. The response is the result of a non-linear Raman interaction with the photons in the parallel polarization[42], which results in the scattering of photons between the two polarization components and modulates the ellipticity of the transmitted light.

The direct output of the detection system is the quadrature distribution. In Fig. 3a,c we see the equilibrium quadrature and its non-equilibrium modulation for two different intensities of the weak probe state. As a function of the time delay, we see that the quadrature distribution changes its width. At the zero delay we have a strong response due to the coherent overlap with the pump pulse, but more importantly we see oscillations at the positive times. The latter are due to intensity, i.e. average photon number, changes of the orthogonal polarization induced by the phonon excitation. Applying the phase-averaged tomography on the quadrature distribution data, we can study the non-equilibrium response of the photon number distribution. The equilibrium distribution in Fig. 3b,d(left) is well fitted by a Poissonian distribution with an average photon number of N = 13.8 and 2.9 photons per pulse, respectively. We observe in Fig. 3b,d(right) that the probability distribution is shifting following the phonon oscillations.

In order to understand how the interaction of the probe with the phonon mode modifies its photon number distribution, we model the light-phonon interaction with a Raman Hamiltonian[42]

$$H_{Ram} = -\sum_\omega \chi \left( a^\dagger_{x,\omega+\Omega} a_{y,\omega} b + a^\dagger_{x,\omega+\Omega} a^\dagger_{y,\omega} b^\dagger + a^\dagger_{y,\omega+\Omega} a_{x,\omega} b + a^\dagger_{y,\omega+\Omega} a^\dagger_{x,\omega} b^\dagger \right) \quad (5)$$

which describes the scattering between the optical polarizations $x, y$ mediated by $\chi = \chi_{xy}$, the non-linear polarizability tensor associated with the $E_T$ phonon. The operator $b$ is the field of the phonon with frequency $\Omega$, while the photon frequency $\omega$ runs over the spectrum of the light pulse. In the specific polarization configuration considered, the pump-probe response is spectrally uniform[42] and for simplicity we will neglect the frequency index $\omega$. We define $x$ as the main polarization axis and $y$ as the weak cross-polarized residual. The input optical coherent states are defined as $a_x|\alpha_x\rangle = \alpha_x|\alpha_x\rangle$, $a_y|\alpha_y\rangle = |\alpha_y|e^{-i\pi/2}|\alpha_y\rangle$, where the factor $\pi/2$ accounts for the phase shift between the two polarization components due to ellipticity.

We are interested in calculating the response of the experimental observable, i.e. the photon number operator $N = a_y^\dagger a_y$ of the weak $y$ polarization. Using a perturbative expansion (with $\tau\chi \ll 1$, where $\tau$ is the interaction duration) we can study the phonon dependent optical response. We calculate (full derivation in the supplementary) the average photon number as a function of the pump-probe delay $\Delta t$

$$\bar{N}(\Delta t) = \alpha_y^2 + 2\tau\chi(|\alpha_x||\alpha_y|)\langle b_{\Delta t}^\dagger + b_{\Delta t}\rangle \qquad (6)$$

and its variance

$$\sigma_N^2(\Delta t) = \bar{N}(\Delta t) \qquad (7)$$
$$+ 4\tau^2\chi^2\alpha_y^2\alpha_x^2\left(\langle b_{\Delta t}^{\dagger\,2}\rangle - \langle b_{\Delta t}^\dagger\rangle^2 + 2(\langle b^\dagger b_{\Delta t}\rangle - \langle b_{\Delta t}^\dagger\rangle\langle b_{\Delta t}\rangle) + \langle b_{\Delta t}^2\rangle - \langle b_{\Delta t}\rangle^2 + 1\right)$$
$$+ \tau^2\chi^2\alpha_y^2\left(2(\langle b^\dagger b_{\Delta t}\rangle - \langle b_{\Delta t}^\dagger\rangle\langle b_{\Delta t}\rangle) + 1\right).$$

The mean-value is oscillating around the initial value at the phonon frequency, ruled by the phonon displacement $q \propto b + b^\dagger$. The variance depends on second order terms of the phonon operator, which means it is in principle sensitive to the phonon statistics. To understand how the variance of the photon number changes according to the phonon properties, we simulate[51] it for different phonon states. We present in Fig. 4 the expected results for a displaced phonon state with coherent (a), thermal (b) and squeezed (c) statistics. We consider the phonon field as a large ensemble of oscillators each with a relatively large excitation amplitude $|\langle b\rangle|=2$ (corresponding to a temperature increase of $\sim$900 K) and set the cross-section ($\tau\chi$) to match the experimentally observed photon number modulation (details on the parameters dependence in the supplementary). The variance of the phonon displacement shows different noise levels and periodicities, peculiar of each phonon state, which are mapped in the optical degrees of freedom by the probe-phonon interaction. We characterize the resulting probe photon distributions in terms of the Mandel parameter $Q$,

$$Q = \frac{\sigma_N^2 - \bar{N}}{\bar{N}} \qquad (8)$$

which quantifies the deviations from a Poissonian photon statistics as a function of the difference between mean ($\bar{N}$) and variance ($\sigma_N^2$). The Mandel parameter is related to the second-order correlation function as $Q = \bar{N}(g^2(0) - 1)$. For an optical coherent state we expect a Poissonian distribution with $Q=0$, with (i.e. $\bar{N} = \sigma_N^2$), while $Q > 0$ or $Q < 0$ describes respectively a super- or sub-Poissonian statistics. The thermal state generates an optical super-Poissonian distribution oscillating at the phonon frequency. The squeezed state produces instead an oscillation at $2\Omega$ around $Q=0$, with its phase determining the shift between the variance modulation and the phonon wave.

To discuss the capabilities and limitations of the proposed technique, we compare model and experiment presenting in Fig. 4d the measured evolution of the photon number mean-value and variance for the $\bar{N} = 2.9$ probe, extracted from the data in Fig.3d. We are able to resolve that both the average photon number and the variance oscillate at the phonon frequency. If we consider the pump-probe response of $Q$, we observe a small super-Poissonian character ($Q > 0$), which is due to classical experimental excess noise. If we correct the Mandel parameter considering the excess noise which affects the detection response, and define $Q_{det}$ (see supplementary), we can explain the deviation from the equilibrium behavior.

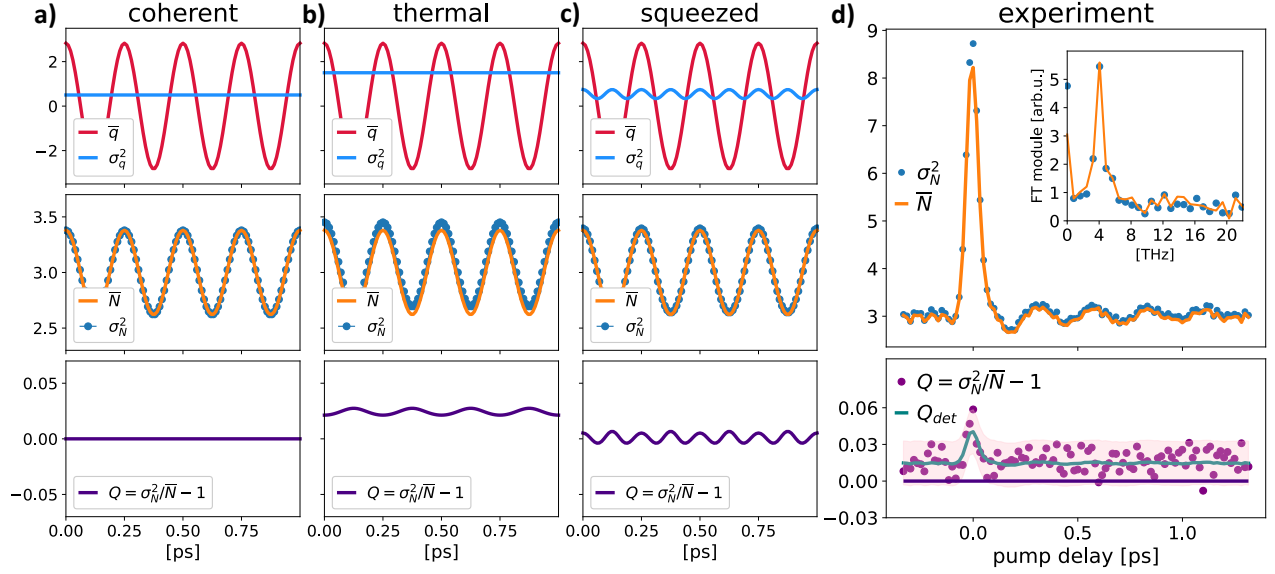

**Figure 4: Phonon dependent evolution of the photon distribution parameters. a,b,c)** We simulate the optical response for a coherent (a), thermal (b) and squeezed (c) phonon states. The top panel compares the average phonon displacement and its variance. The middle panel reports the optical response (average and variance of the photon number). The Mandel parameter Q (bottom) quantifies the deviations from the coherent state Poissonian statistics. **d)** The experimental mean photon number and variance oscillate at the 4 THz phonon frequency, as shown by the Fourier Transform analysis of the positive times (insert). The data are consistent with the detector response $Q_{det}$ (green line), which describes a Poissonian behavior (Q = 0) corrected considering the intensity dependent classical excess noise (see supplementary). The pink area accounts for the error calculated as standard deviation of repeated measurements of Q (2σ).

## Discussion

The experiment and the calculations performed in this work validate a new methodology to perform ultrafast time-resolved quantum spectroscopy experiments, which is suitable to reveal signatures of the quantum nature of light-matter interaction in the phase-averaged optical statistical degrees of freedom. We proved the capability to measure non-equilibrium changes in the photon number distribution with ultrafast resolution and we showed theoretically that the material fluctuations can qualitatively perturb the optical statistics. Our measurement on quartz sets a bound on the amount of squeezing or thermal excitations present in the system.

The model predicts qualitative changes in the optical response as a function of the phonon fluctuations, which are quantitatively comparable to the detection noise. Using a Bose-Einstein distribution ($n = 1/(e^{\hbar\Omega/k_B T} - 1)$) we can expect an equilibrium average occupation of 0.7 levels for the 4 THz (16.5 meV) phonon at 300 K. The state reported in Fig.4b is simulated with a thermal occupation of 1, which should be observable with the current experimental conditions. The result suggests that the non-equilibrium state is well-described by a coherent excitation, without an appreciable injection of incoherent thermal population. To reveal possible modifications due to the quantum fluctuations in this system a higher acquisition statistic is required. This issue can be overcome using a laser source with higher repetition rate. The current setup could successfully reveal quantum effects in systems with a more pronounced non-Poissonian character or stronger light-matter couplings.

This method establishes a direct connection between the quantum fluctuations of a material and the statistical properties of the electric fields, thus potentially constituting an indicator of "quantumness". Since the variance of the photon number encodes the variance of microscopic observables, this quantity could be used as tool to witness entanglement in many-body systems of interest. The variance of a quantum operator is the sum of a thermal/incoherent part that can be separated from the coherent/quantum part, the quantum variance[52], which provides a lower bound for fundamental quantum estimators, such as the quantum Fisher information (QFI). The QFI is a witness of multipartite entanglement[53], which can be quantified in solids by performing a full integration over the energy spectrum of mean-value dynamical susceptibilities[54,55]. Our method constitutes a direct statistical approach to probe quantum fluctuations of a macroscopic solid and to bound the entanglement associated with specific degrees of freedom.

In addition to quantum states, the method can allow us to explore statistical effects also in the classical regime. Unlike the uniform coherent phonon excitation in quartz, strong fluctuations can be present in an inhomogeneous system, where the state is a statistical mixture of distinguishable oscillators. For instance, if the dephasing between different oscillators is faster than the population decay, we can expect the evolution of the system into an incoherent state with super-Poissonian statistics. The classical statistical response can then be used in this framework to distinguish the response of the dynamics of population and coherence[56].

In prospective, the proposed method opens the way to a new typology of quantum spectroscopy based on the study of the statistical response of weak coherent laser pulses, which can be useful to design ultrafast photonic quantum devices and reveal insight on the non-equilibrium dynamics of fluctuations in complex materials.

## Methods

The sample is an α-quartz crystal, with 1 mm thickness.

The laser pulses are obtained from a pulsed source+OPA system (Pharos+Orpheus, LightConversion). The laser source delivers 1026 nm, 120 fs pulses at 1 kHz repetition rate and we set the outputs of the OPA (Signal and Idler) such that the SH of the Idler (1540 nm, 0.805 eV, <100 fs) is resonant to the Signal (770 nm, 1.61 eV, <50 fs).

The experiment is performed in transmission. The equilibrium ratio between parallel and residual intensities due to quartz birefringence is about 100. The fluences on the quartz sample are 4 mJ/cm2 for the pump and 2-8 nJ/cm2 for the probe. The employed LO has parallel polarization with respect to the detected probe and the full intensity at the detection beam-splitter is about 1 pJ ($10^9$ photons/pulse). The reported data are acquired with a spectrally shaped LO with a narrow 0.5 meV bandwidth, centered at 1.62 eV ($10^7$ photons/pulse). We collect trains of 4000 pulses per delay point and scan the pump-probe trace 60 times.

The phonon statistics and evolution are simulated using the software tools from the QuTiP package[51].

## Acknowledgments

This work was mainly supported by the European Research Council through the project INCEPT (grant Agreement No. 677488). DF, EMR, AM, and GJ acknowledge the support of the Gordon and Betty Moore foundation through the grant (CENTQC).

# Supplementary material

**A. Uniform random sampling of the quadrature phase**

**B. Multimode probe characterization**

**C. Quantum shot-noise characterization**

**D. Quantum model for Impulsive Stimulated Raman Scattering**

**E. Statistical properties of the phonon state**

**F. Maximum likelihood phase-averaged tomography algorithm**

## A  Uniform random sampling of the quadrature phase

We study the randomness of the Carrier-Envelope Phase (CEP) statistical distribution and its key role in ensuring a reliable measurement of the phase-averaged quadrature distribution of the optical probe state. The intrinsic CEP fluctuations exploited in our technique are perfectly uniformly distributed and uncorrelated, granting a homogeneous sampling of the phase of the quadrature which is insensitive to the phase instabilities affecting the interferometer.

We demonstrate the randomness of the CEP adopted in our experiment by analyzing the quadrature values recorded individually for a train of consecutive probe pulses. In Fig.S1a, we observe that the measurements of the optical quadrature, $X = \frac{1}{\sqrt{2}}(ae^{i\phi} + a^\dagger e^{-i\phi})$, span the entire amplitude range without any distinguishable trend between subsequent pulses. We note that the increased occurrence of measurements at the extremes of the range of $X$ is consistent with the shape of the quadrature distributions presented in the main text (Fig.2c), which is a consequence of the projection of the ring-like Wigner distribution of a phase-averaged coherent state. The features of the measured dataset are consistent with those of a simulated dataset (Fig.S1b), where we assume a uniform random distribution for $\phi$ between 0 and $2\pi$, and add a Gaussian fluctuation $\delta$ ($\sigma_\delta^2 = 1/2$) to the quadrature to account for the vacuum fluctuations. The agreement between experiment and simulation also holds if we analyze the correlation between successive pulses by plotting the value recorded for the $(i+1)^\text{th}$ pulse as a function of the preceding one. As shown in Fig.S1c,d, no trend is revealed, meaning that correlations between successive pulses are absent. In Fig.S1e, we quantitatively verify the absence of correlations in a train of subsequent measurements calculating the Pearson correlation coefficient as a function of the distance between pulses, which trivially equals 1 for the auto-correlation and reads zero for any other following pulse.

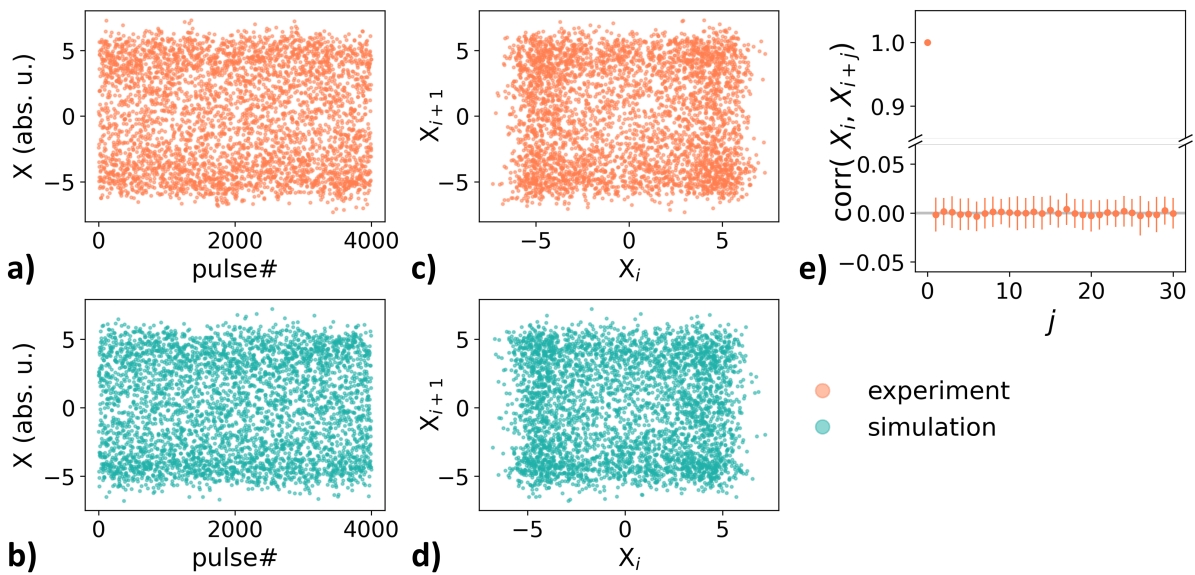

Figure S1: **Characterization of Carrier-Envelope Phase randomness, comparing experimental data with simulations using a uniformly random phase.** (**a**) Experimental and (**b**) simulated quadrature values for consecutive single-pulse acquisitions. (**c, d**) Lack of correlation between individual quadrature measurements of two adjacent pulses. (**e**) Correlation coefficient calculated between pairs of quadrature acquisitions in a sequence of subsequent pulses.

Conducting phase-averaged studies using the uniform and uncorrelated phase sampling resulting from CEP instability is the most reliable and time-efficient approach. In Fig.S2, we report simulated results comparing the phase-randomized method with other schemes that use a controllable phase (for instance, piezoelectric translators to finely tune the interferometer path length). The random CEP scheme has three main advantages: i) the phases are sampled in a perfectly uniform way, ii) the time to switch between different phases is as fast as the repetition rate of the employed pulses, and iii) it is insensitive to environmental instabilities which can produce both slow drifts and sudden jumps in the phase recorded by the interferometer. A controlled sampling of a reduced number of phases (Fig.S2b) would introduce artifacts in the quadrature distribution due to finite sampling and would be slowed down by the dead times spent tuning the phase during the acquisition scans. A continuously varying linear phase scan (Fig.S2c) would be a better strategy to generate a correct phase-averaged quadrature distribution, but any systematic deviation from perfect linearity (d) or stochastic fluctuation (e) would distort the final result. Thus, phase-controlled methods can be considered as alternatives for implementing approximately exact phase-averaged tomography, but the ability to exploit the properties of CEP instability makes the phase-randomized method the optimal choice for phase-averaged protocols.

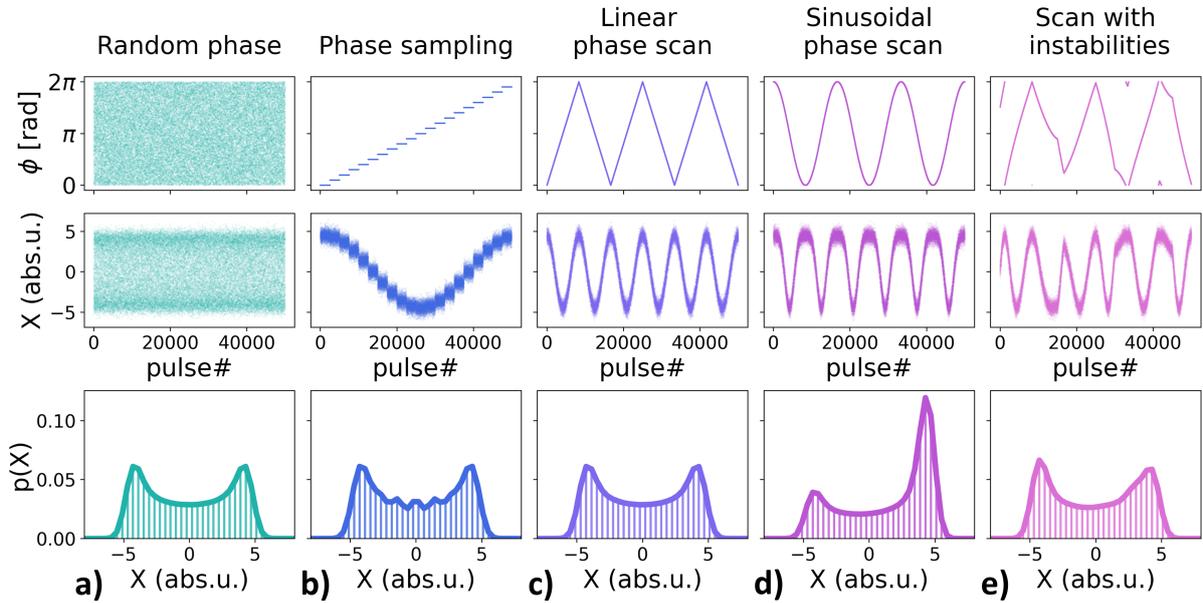

Figure S2: **Simulated results for different strategies to implement a phase-averaged detection of coherent state.** A specific sequence of phases set at each individual pulse realization (top) produces a set of optical quadrature acquisitions (center) corresponding to different phase-integrated statistical distributions (bottom). The random phase configuration (**a**) and a perfectly linear phase scan (**c**) both produce a perfect phase-averaged distribution. However, finite sampling effects (**b**) or deviations from the ideal linear scan (**d,e**) introduce artifacts in the quadrature distribution.

## B   Multimode probe characterization

We characterize the multimode spectrum of the ultrashort probe pulse. The frequency-resolved detection is obtained by intereference with a narrow frequency Local Oscillator (bandwidth 0.5 meV), whose spectral content is selected by a pulse shaper [26]. In particular, in our setup we use a Liquid Crystal Spatial Light Modulator (LC-SLM) in a diffraction-based scheme (Fig.S3a). In Fig.S3b we show the photon number distribution obtained as a function of the spectral frequency for some selected frequency modes. We observe as the distribution shifts following the Gaussian profile of the frequency spectrum. From the full distribution we detail the mean value and variance spectra. Mean and variance are superimposed as expected for coherent states with Poissonian statistics.

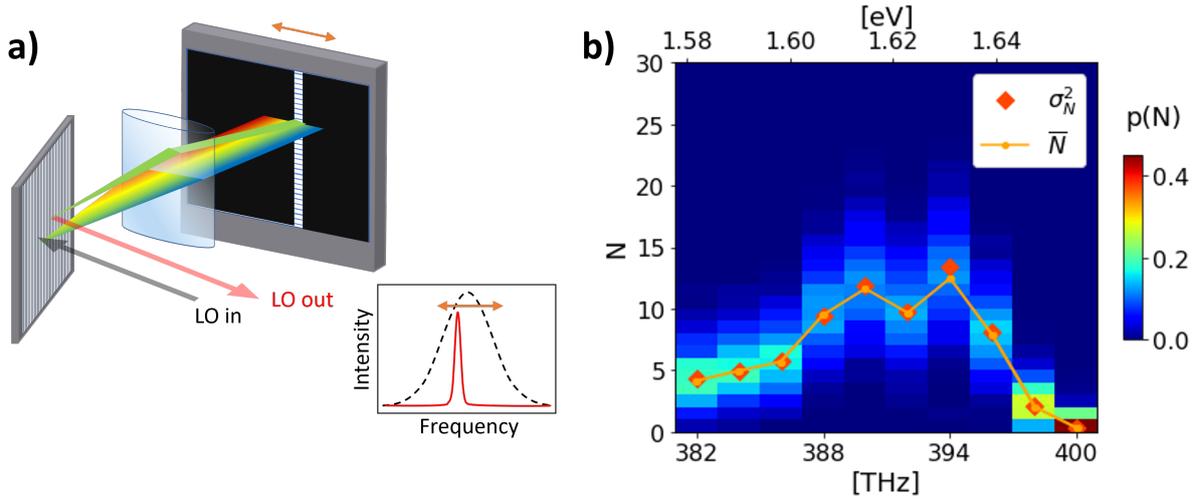

Figure S3: **Equilibrium characterization of the multimode probe spectrum. a)** Diffraction-based pulse shaping scheme. A programmable LC-SLM is adopted to select the Local Oscillator spectral content and implement a frequency-resolved detection. **b)** We report the photon number distribution measured for different probe frequency components. From the full distribution map we calculate the spectra for the mean photon number (orange circles) and the relative variance (red diamonds).

## C   Quantum shot-noise characterization

Our experimental scheme relies on the rejection of excess noise (i.e. classical experimental fluctuations) which adds to the quantum shot-noise level. In an ideal setting, all the classical fluctuations are perfectly cancelled by the balanced differential detection. However, some residual unbalance is unavoidable, e.g. the one induced by the homodyne interference. In Fig.S4 we characterize the noise as a function of the probe intensity. The classical fluctuations are usually quadratic in the photon number, while the quantum ones are linear. As a consequence of this, we can reduce the relative contribution of classical noise using weak coherent states. From the data in Fig.S4, we can fit the experimental noise with a quadratic trend and compare classical and quantum contributions ($\sigma_N^{2\,(det)} = N + p_2 N^2 + p_1 N + p_0$). The parameters $p_1$ and $p_0$ are negligibly small and we estimate that for a mean photon number $< 10$ we have a classical to quantum noise ratio $< 2\%$.

The excess noise gives a super-Poissonian statistics and a positive value of the Mandel parameter. Basing on the fit on (a), we can calculate the expected Mandel parameter $Q_{det} = \sigma_N^{2\,(det)}/N - 1 \simeq p_2 N$ and employ it to reference changes in the non-equilibrium dynamics.

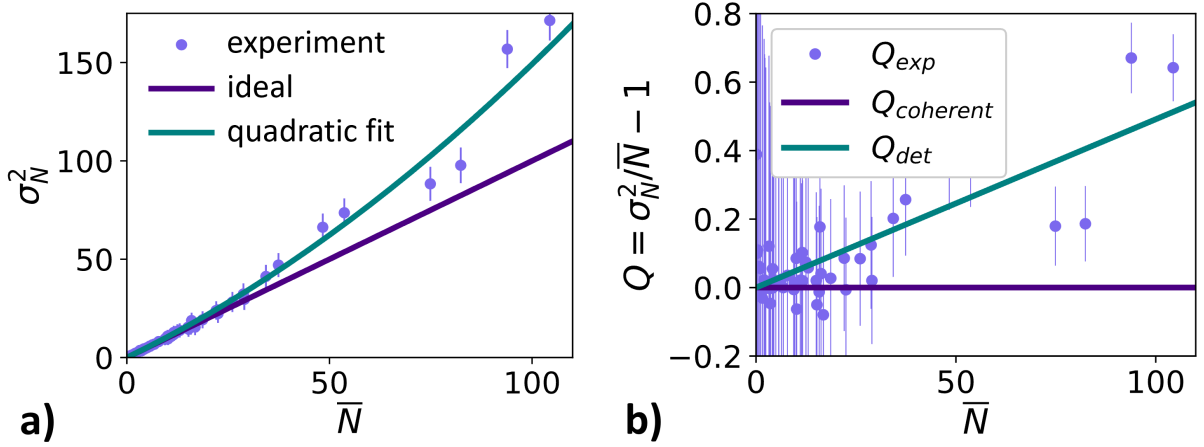

Figure S4: **Excess noise characterization at equilibrium as a function of the mean photon number.** a) The measurement of the photon number variance progressively deviates from the ideal Poissonian behavior with increasing mean photon number (error bars represent Poissonian noise, $1\sigma$). The contribution of the classical fluctuations is referenced using a quadratic polynomial fit (green line). b) Calculation of the Mandel parameter $Q$. $Q_{det}$ describes the coherent state response taking into account the detection excess noise.

## D  Quantum model for Impulsive Stimulated Raman Scattering

We study how different phonon statistical properties affect the optical response in a pump-probe experiment using the model presented in the article *Quantum model for Impulsive Stimulated Raman Scattering*, J. Phys. B: At. Mol. Opt. Phys. 52, 145502 (2019) [42], and in chap. 6-7 of F. Glerean's PhD Thesis (https://hdl.handle.net/11368/2988327).

The light-phonon interaction under consideration is a Raman interaction, which can be modeled modeled with an Hamiltonian of the form

$$H_{Ram} = -\sum_{\lambda,\lambda'} \chi^{(1)}_{\lambda,\lambda'} \sum_j \left[ \left(a^\dagger_{\lambda j} a_{\lambda' j+\frac{\Omega}{\delta}}\right) b^\dagger + \left(a_{\lambda j} a^\dagger_{\lambda j+\frac{\Omega}{\delta}}\right) b \right] \quad (1)$$

where $a_{\lambda,j}$ and $b$ represent respectively the photon and phonon fields, $j$ labels the frequency component of the multi-mode optical pulse ($\omega = \omega_0 + \delta j$) and $\lambda$ is the polarization index.

The non-linear polarizability tensor $\chi^{(1)}_{\lambda,\lambda'}$ regulates the strength of the interaction and the symmetry of the phonon mode. We analyze the response of the $E^T$ symmetry mode, which has off-diagonal susceptibility terms, resulting in photon scattering among orthogonal polarizations.

$$H_{Ram} = -\chi^{(1)}_{x,y} \sum_j \left[ \left(a^\dagger_{x,j} a_{y,j+\frac{\Omega}{\delta}}\right) b^\dagger + \left(a_{x,j} a^\dagger_{y,j+\frac{\Omega}{\delta}}\right) b + \left(a^\dagger_{y,j} a_{x,j+\frac{\Omega}{\delta}}\right) b^\dagger + \left(a_{y,j} a^\dagger_{x,j+\frac{\Omega}{\delta}}\right) b \right] \quad (2)$$

In the experiment, the probe pulse is mainly polarized along $x$, and we measure the modulation of the phonon induced scattering in the residual orthogonal component $y$ with a small number of photons per pulse. We calculate the effects of the interaction hamiltonian in the optical field considering the evolution of the operators up to second order in the evolution parameter $\tau \chi^{(1)}$ ($\tau$ is the interaction time) as

$$a'_{yj} = a_{yj} + i\tau[H_{Ram}, a_{yj}] - \frac{\tau^2}{2}[H_{Ram},[H_{Ram}, a_{yj}]] \quad (3)$$

with

$$[H_{Ram}, a_{yj}] = +\chi^{(1)}(a_{xj+\frac{\Omega}{\delta}} b^\dagger + a_{xj-\frac{\Omega}{\delta}} b) \quad (4)$$

and

$$[H_{Ram},[H_{Ram},a_{yj}]] = +(\chi^{(1)})^2\Big(a_{yj+\frac{2\Omega}{\delta}}b^\dagger b^\dagger + a_{yj}b^\dagger b + a_{yj}bb^\dagger + a_{yj-\frac{2\Omega}{\delta}}bb \quad (5)$$
$$+ \sum_j(-a_{xj+\frac{\Omega}{\delta}}a_{xj}a^\dagger_{yj+\frac{\Omega}{\delta}} - a_{xj+\frac{\Omega}{\delta}}a^\dagger_{xj+\frac{\Omega}{\delta}}a_{yj} \quad (6)$$
$$+ a_{xj-\frac{\Omega}{\delta}}a^\dagger_{xj}a_{yj+\frac{\Omega}{\delta}} + a_{xj-\frac{\Omega}{\delta}}a_{xj+\frac{\Omega}{\delta}}a^\dagger_{yj}\Big). \quad (7)$$

The state on which the operators act is a multimode coherent light state combined with a statistical mixture of coherent phonon states. In the density operator formalism we can describe it as

$$\rho = \rho_{light} \otimes \rho_{phonon} = (|\alpha\rangle\langle\alpha|) \otimes (\sum_m p_m |\beta_m\rangle\langle\beta_m|) \quad (8)$$

where $|\alpha\rangle = \otimes_j |\alpha_j\rangle$ (with $a_j|\alpha_j\rangle = \alpha_j|\alpha_j\rangle$) is the multimode state built as tensor product of the individual light modes, while the phonon state is a mixed state describing the statistical ensemble of identical phonon oscillators, where each state $|\beta_m\rangle$ occurs with probability $p_m$.

Our goal is to study the fluctuations of the system. We can calculate the variance of the generic operator O as

$$\sigma_O^2 = \langle O^2 \rangle - \langle O \rangle^2 = Tr(\rho O^2) - Tr(\rho O)^2 \quad (9)$$

In order to test the possible results of the experiment we focus on the probe intensity (i.e. the photon number) $N_{yj} = a^\dagger_{yj}a_{yj}$. The probe output operator after the interaction with the sample is precisely

$$N'_{yj} = (a'_{yj})^\dagger a'_{yj} \quad (10)$$

which up to second order explicitly reads

$$N'_{yj} = (a_{yj})^\dagger a_{yj} + i\tau(a^\dagger_{yj}[H,a_{yj}] - [H,a_{yj}]^\dagger a_{yj}) + \quad (11)$$
$$+ \tau^2[H,a_{yj}]^\dagger[H,a_{yj}] \quad (12)$$
$$- \frac{\tau^2}{2}(a^\dagger_{yj}[H,[H,a_{yj}]] + [H,[H,a_{yj}]]^\dagger a_{yj}) \quad (13)$$

from which we can compute

$$\langle N' \rangle = Tr(\rho N') \quad (14)$$

and

$$\langle N'^2 \rangle = Tr(\rho N'^2). \quad (15)$$

The expectation value of the probe intensity:

$$\langle N'_{yj} \rangle = \langle (a_{yj})^\dagger a_{yj} \rangle + i\tau \langle a^\dagger_{yj}[H,a_{yj}] - [H,a_{yj}]^\dagger a_{yj} \rangle \quad (16)$$
$$+ \langle \tau^2[H,a_{yj}]^\dagger[H,a_{yj}] \rangle \quad (17)$$
$$- \frac{\tau^2}{2} \langle a^\dagger_{yj}[H,[H,a_{yj}]] + [H,[H,a_{yj}]]^\dagger a_{yj} \rangle \quad (18)$$

at first order can be expressed as

$$\langle N'_{yj} \rangle = \langle a^\dagger_{yj} a_{yj} \rangle + \tag{19}$$
$$+ i\tau\chi\Big( \langle b^\dagger \rangle \big( \langle a^\dagger_{yj} \rangle \langle a_{xj+\frac{\Omega}{\delta}} \rangle + \langle a_{yj} \rangle \langle a^\dagger_{xj-\frac{\Omega}{\delta}} \rangle \big) \tag{20}$$
$$- \langle b \rangle \big( \langle a^\dagger_{yj} \rangle \langle a_{xj-\frac{\Omega}{\delta}} \rangle + \langle a_{yj} \rangle \langle a^\dagger_{xj+\frac{\Omega}{\delta}} \rangle \big) \Big) \tag{21}$$
$$\tag{22}$$

The calculation of the squared intensity operator up to second order reads:

$$N^{2'}_{yj} = a^\dagger_{yj} a_{yj} a^\dagger_{yj} a_{yj} + \tag{23}$$
$$+ i\tau\Big( a^\dagger_{yj} a_{yj} a^\dagger_{yj} [H, a_{yj}] - a^\dagger_{yj} a_{yj} [H, a_{yj}]^\dagger a_{yj} \tag{24}$$
$$+ a^\dagger_{yj}[H, a_{yj}] a^\dagger_{yj} a_{yj} - [H, a_{yj}]^\dagger a_{yj} a^\dagger_{yj} a_{yj} \Big) \tag{25}$$
$$+ \tau^2\Big( - a^\dagger_{yj}[H, a_{yj}] a^\dagger_{yj} [H, a_{yj}] + a^\dagger_{yj}[H, a_{yj}][H, a_{yj}]^\dagger a_{yj} \tag{26}$$
$$+ [H, a_{yj}]^\dagger a_{yj} a^\dagger_{yj} [H, a_{yj}] - [H, a_{yj}]^\dagger a_{yj} [H, a_{yj}]^\dagger a_{yj} \tag{27}$$
$$+ a^\dagger_{yj} a_{yj} [H, a_{yj}]^\dagger [H, a_{yj}] + [H, a_{yj}]^\dagger [H, a_{yj}] a^\dagger_{yj} a_{yj} \tag{28}$$
$$- \frac{1}{2}(a^\dagger_{yj} a_{yj} a^\dagger_{yj} [H, [H, a_{yj}]] + a^\dagger_{yj} a_{yj} [H, [H, a_{yj}]]^\dagger a_{yj}) \Big) \tag{29}$$

If we use the previous result to calculate the variance, taking into account the commutation relation for $[a_{yj}, a^\dagger_{yj}] = 1$ we obtain

$$\sigma^2_{N'_{yj}} = \langle N^{2'}_{yj} \rangle - \langle N'_{yj} \rangle^2 = \langle N^{2'}_{yj} \rangle + \tag{30}$$
$$+ \tau^2\Big( - \langle a^\dagger_{yj} a^\dagger_{yj} \rangle (\langle [H, a_{yj}][H, a_{yj}] \rangle - \langle [H, a_{yj}] \rangle \langle [H, a_{yj}] \rangle) \tag{31}$$
$$+ \langle a^\dagger_{yj} a_{yj} \rangle (\langle [H, a_{yj}]^\dagger [H, a_{yj}] \rangle - \langle [H, a_{yj}]^\dagger \rangle \langle [H, a_{yj}] \rangle) + h.c.\Big) \tag{32}$$

where the first line is the shot noise proportional to the photon number, while the other factors at second order depend on the phonon statistics.

In order to reach a tractable final expression, we underline that we are working with birefringent quartz and that orthogonal modes are out of phase because of the equilibrium refraction which generates elliptical polarization. The ellipticity can be taken into account setting $\phi_{\alpha_x} = 0$ and $\phi_{\alpha_y} = \pi/2$. This makes $\langle a^\dagger_{yj} a^\dagger_{yj} \rangle = |\alpha_{yj}|^2 e^{-i\pi} = -|\alpha_{yj}|^2 = - \langle a^\dagger_{yj} a_{yj} \rangle$, which is crucial to avoid mutual cancellation of additional variance terms.

We explicit the $a_{xj}$ and $b$ fields, taking into account the additional terms resulting from their commutation relations. To simplify and write the final result for the variance in a more compact way, we also neglect the spectral dependence on the light amplitude $\alpha_{xj} = \alpha_x$.

$$\sigma^2_{N'_y} = \langle N^{2'}_y \rangle + \tag{33}$$
$$+ 4\tau^2 \chi^2 \alpha_y^2 \alpha_x^2 \Big( \langle (b^\dagger + b)^2 \rangle - \langle b^\dagger + b \rangle^2 \Big) \tag{34}$$
$$+ \tau^2 \chi^2 \alpha_y^2 \Big( 2(\langle b^\dagger b \rangle - \langle b^\dagger \rangle \langle b \rangle) + 1 \Big). \tag{35}$$

To quantify how the quantum statistical response scales when describing a macroscopic field we consider the dependence on the size of an ensemble of $M$ identical and independent probe phonon oscillators, for which we have a total mean phonon number

$$N_{phonons} = M \langle b^\dagger b \rangle. \tag{36}$$

Basing on this, we scale the field operator as $b \longrightarrow \sqrt{M} b$ in the model predictions, which gives

$$\langle N'_y \rangle = \alpha_y^2 + \tau\chi|\alpha_y||\alpha_x|\sqrt{M} \langle b^\dagger + b \rangle \tag{37}$$

and

$$\sigma^2_{N'_y} = \langle N'_y \rangle + \tag{38}$$
$$+ \; 4\tau^2\chi^2\alpha_y^2\alpha_x^2 \Big( M\big( \langle b^{\dagger 2} \rangle - \langle b^\dagger \rangle^2 + 2(\langle b^\dagger b \rangle - \langle b^\dagger \rangle \langle b \rangle) + \langle b^2 \rangle - \langle b \rangle^2 \big) + 1 \Big) \tag{39}$$
$$+ \; \tau^2\chi^2\alpha_y^2 \Big( 2M(\langle b^\dagger b \rangle - \langle b^\dagger \rangle \langle b \rangle) + 1 \Big). \tag{40}$$

## E  Statistical properties of the phonon state

The optical photon number distribution is sensitive to the statistical properties of the phonon state. We study how different phonon states are mapped in the optical observables using numerical simulations [51] of the phonon statistics for some prototypical states of a quantum harmonic oscillator. We present in Fig.S5 the Wigner distribution describing in the phase space the coherent, thermal and squeezed states employed in the simulations. The pump excitation displaces the system from the origin of the phase space. We consider a displacement operator $D(\beta) = \exp(\beta b^\dagger - \beta^* b)$ with $\beta = 2$, The differences between the states are in their statistical phase space distributions. The coherent state has Heisenberg (vacuum) limited uncertainty in both position and momentum ($\sigma_q^2 = \sigma_p^2 = 1/2$). The thermal state has a larger distribution, and we set its width corresponding to a thermal population of 1 phonon. The Bose-Einstein distribution ($\overline{n} = 1/(e^{\hbar\omega/k_B T} - 1)$) predicts an occupation of about 0.7 phonons at 300 K for the 4 THz mode studied in quartz. The squeezed state distributes the minimal Heisenberg uncertainty anisotropically between position and momentum. We considered the action of a squeezing operator $S(\zeta) = \exp[1/2(\zeta^* b^2 - \zeta b^{\dagger 2})]$, with a real squeezing parameter $\zeta = -0.2$. In general, $\zeta$ is a complex parameter, the phase of which defines the direction of the squeezing in the phase space, while the amplitude controls the magnitude of the effect. The example in Fig.S5c, with real $\zeta$, models squeezing aligned along the phase space axes. The negative sign results in an increase of the variance at the crest and a decrease at the nodes of the position and momentum operators. A different phase of $\zeta$ would shift this redistribution of fluctuations with respect to the phase of the phonon oscillation, modifying the phase but not the amplitude of the changes observed in the phonon statistics. The time-dependent evolution of the phonon is described by a rotation of the state in the phase space around the origin. Formally, we calculate it considering the unitary evolution $U = e^{-iHt}$ ruled by the phonon hamiltonian $H = \Omega b^\dagger b$.

We calculate the optical response inserting the phonon mean displacement and variance in the expression obtained with the model for the Raman interaction. As in the main text, the equilibrium mean value of the probe photons is $N_y = 3$. We set the cross-section parameters $\tau\chi$ so that the product $|\beta|\tau\chi\sqrt{M}$ matches the amplitude of the experimental response.

In Fig.S6 we investigate the quantitative dependence on the simulation parameters. We study the dependence on the probe intensity, phonon amplitude and number of probed phonon oscillators. We use as observable the maximum of the optical Mandel parameter over a phonon oscillation. In Fig.S6a we observe that the deviation from the Poissonian behavior increases with the probe intensity, but quickly saturates. For high photon numbers the optical statistics is dominated by the classical detection noise, estimated in appendix C. We understand that although compromising the absolute amplitude of the pump-probe mean-value response, performing the experiment in the few-photon regime is crucial to highlight the contribution of the quantum fluctuations and avoid the one related to the classical noise. In Fig.S6b, we vary $|\beta|$ up to 6 (corresponding to 7000 K for the 4 THz $E_T$ phonon in quartz) keeping the product $|\beta|\tau\chi$ constant. We observe that the modulation of the statistics, measured by $Q$, decreases as we increase $\beta$ (i.e. decreasing the cross-section). This indicates that the quantum effects of the light-photon interaction become less important with increasing phonon amplitude or smaller susceptibility.

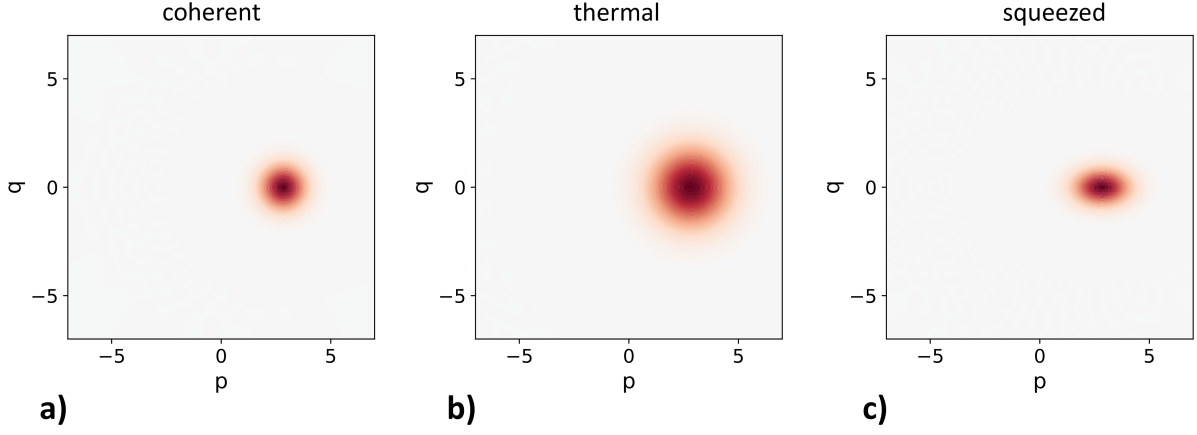

**Figure S5: Representation of the phonon state in the phase space.** Wigner distributions describing in the phonon position-momentum phase space the coherent (**a**), thermal (**b**) and squeezed states (**c**) employed in the simulations.

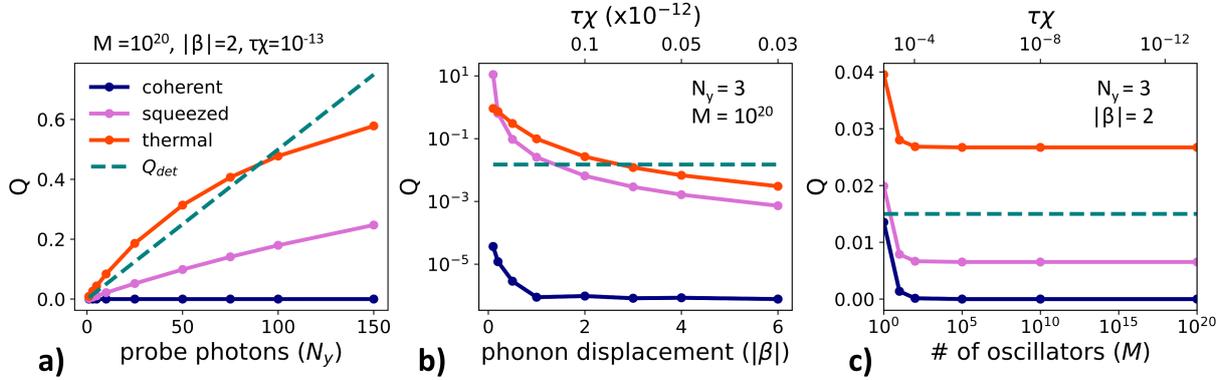

**Figure S6: Dependence of the Mandel parameter maximum on the simulation parameters for the different phonon states. a)** The deviation from the Poissonian behavior increases with the number of probe photons, but is soon dominated by the increase in the classical detection noise ($Q_{det}$). **b)** Sets of parameters with the same change in the average photon number give bigger statistical effects for smaller phonon displacements, which correspond to larger non-linear susceptibilities. **c)** The statistical effects are not averaged away when measuring large ensembles of oscillators. Potential quantum effects related to the interaction with the phonon vacuum emerge close to the single system regime.

The results in Fig.S6c show that the amplitude of the statistical effect tend to a well defined value in the limit of a large number of oscillators, without strong dependence on the size of the probed sample. This is important because it reveals that the statistical effects can be accessed probing macroscopic samples. Interestingly, the $Q$ factor increases for all the phonon distributions close to 1. The latter effect is due to the photon interaction with the phonon vacuum state, which is quantified by the unit factors generated by the commutation relation for the phonon operators $[b, b^\dagger] = 1$. The detection of this latter effect requires though the challenging capability to resolve the response from a microscopic unit.

## F  Maximum likelihood phase-averaged tomography algorithm

We report on the algorithm employed to perform the phase-averaged tomography procedure. Our goal is to reconstruct the photon number distribution $p(N)$ of the probe state from the measurement of the quadrature $X$, in particular from the data of its phase-averaged (PHAV) statistical distribution $f(X)$.

The photon number distribution is expressed in terms of the density operator as the diagonal elements of the density matrix in the Fock basis

$$p(N) = \langle N | \hat{\rho} | N \rangle. \tag{41}$$

From the practical point of view, we consider a truncated Fock space with $N < N_{max}$. The numerical limitation of the employed calculator is $N_{max} = 150$.

In order to calculate these terms we use an iterative algorithm which retrieves the density operator which maximizes the probability to obtain the measured data. According to the Maximum Likelihood approach [39,41], we can retrieve iteratively an estimation of the density operator with the following relation

$$\hat{\rho}^{k+1} = \mathcal{N}[\hat{R}(\hat{\rho}^k)\hat{\rho}^k \hat{R}(\hat{\rho}^k)], \tag{42}$$

where $\mathcal{N}$ accounts for normalization.

The start density operator, $\hat{\rho}^0$, is arbitrary and we set it as the normalized identity operator. The evolution is then obtained calculating the effect of the $\hat{R}$ operator. The $\hat{R}$ operator reads

$$\hat{R}(\hat{\rho}^k) = \int_{-\infty}^{+\infty} dX f(X) \frac{\hat{\Pi}(X)}{\mathrm{Tr}[\hat{\Pi}(X)\hat{\rho}^k]}. \tag{43}$$

The $\hat{\Pi}$ operator describes the measurement process. The phase-resolved homodyne measurement is described by the projector $\hat{\Pi}(\theta, X)$, which is expressed in the Fock basis as

$$\langle m | \hat{\Pi}(\theta, X) | N \rangle = \langle m | \theta, X \rangle \langle \theta, X | N \rangle \tag{44}$$

where

$$\langle N | \theta, X \rangle = e^{in\theta} \left(\frac{2}{\pi}\right)^{\frac{1}{4}} \frac{H_N(\sqrt{2}X)}{\sqrt{2^N N!}} \exp(-X^2). \tag{45}$$

In our specific case, the PHAV projector reads

$$\hat{\Pi}(X) = \frac{1}{2\pi} \int_0^{2\pi} d\theta \, \hat{\Pi}(\theta, X) \tag{46}$$

for which the previous expressions reduce to

$$\langle m | \hat{\Pi}(X) | N \rangle = \delta_{m,N} |\langle N | X \rangle|^2, \tag{47}$$

with

$$|\langle N | X \rangle|^2 = \left(\frac{2}{\pi}\right)^{\frac{1}{2}} \frac{(H_N(X))^2}{2^N N!} \exp\left(-\frac{X^2}{2}\right)^2. \tag{48}$$

when the normalization $X = \frac{1}{\sqrt{2}}(\hat{a} + \hat{a}^\dagger)$ is employed.

The matrix elements of the $\hat{\Pi}(X)$ and $\hat{\rho}^k$ operators are the fundamental blocks used in calculating the output. The PHAV setting doesn't allow us to reconstruct $\hat{\rho}$ completely, but it is enough to calculate the diagonal elements describing the photon number distribution $\langle N| \hat{\rho}^{k+1} |N\rangle$.

The not-normalized output is expressed by

$$\langle N| \hat{\rho}^{k+1} |N\rangle = \sum_{l,m} \langle N| \hat{R}(\hat{\rho}^k) |l\rangle \langle l| \hat{\rho}^k |m\rangle \langle m| \hat{R}(\hat{\rho}^k) |N\rangle \tag{49}$$

$$= \langle N| \hat{\rho}^k |N\rangle \left( \int_{-\infty}^{+\infty} dX\, f(X) \frac{\langle N| \hat{\Pi}(X) |N\rangle}{\text{Tr}[\hat{\Pi}(X)\hat{\rho}^k]} \right)^2 \tag{50}$$

where

$$\text{Tr}[\hat{\Pi}(X)\hat{\rho}^k] = \sum_{l,m} \langle m| \hat{\Pi}(X) |l\rangle \langle l| \hat{\rho}^k |m\rangle = \sum_{m} |\langle m|X\rangle|^2 \langle m| \hat{\rho}^k |m\rangle, \tag{51}$$

in which we used eq. 47.

The calculation is then completed normalizing the output as

$$\langle N| \hat{\rho}_{norm}^{k+1} |N\rangle = \frac{\langle N| \hat{\rho}^{k+1} |N\rangle}{\sum_m \langle m| \hat{\rho}^{k+1} |m\rangle}. \tag{52}$$

The results shown in the present work are obtained running the algorithm for 100 iterations.